\documentclass[%
 aip,
 amsmath,amssymb,
 reprint,%
]{revtex4-2}

\usepackage{graphicx}% Include figure files
\usepackage{dcolumn}% Align table columns on decimal point
\usepackage{bm}% bold math
\usepackage[utf8]{inputenc}
\usepackage[T1]{fontenc}
\usepackage{mathptmx}
\usepackage{etoolbox}

\makeatletter
\def\@email#1#2{%
 \endgroup
 \patchcmd{\titleblock@produce}
  {\frontmatter@RRAPformat}
  {\frontmatter@RRAPformat{\produce@RRAP{*#1\href{mailto:#2}{#2}}}\frontmatter@RRAPformat}
  {}{}
}%
\makeatother
\begin{document}

\preprint{AIP/123-QED}

\title{Two-level System Loss: Significant not only at Millikelvin }
% Force line breaks with \\
\author{W. Shan}
%\thanks{}
\affiliation{National Astronomical Observatory of Japan, Mitaka, Tokyo, 181-8588, Japan}

\author{S. Ezaki}
\affiliation{National Astronomical Observatory of Japan, Mitaka, Tokyo, 181-8588, Japan}

\date{\today}% It is always \today, today,

\begin{abstract}
Tow-level system (TLS) loss in amorphous dielectric materials has been intensively studied at millikelvin temperatures due to its impact on superconducting qubit devices and incoherent detectors. However, the significance of TLS loss in superconducting transmission lines at liquid helium temperatures remains unclear. This study investigates TLS loss in amorphous $\rm{SiO_2}$ at liquid helium temperatures (about 4 K) within a frequency range of 130-170 GHz, using  niobium microstrip and coplanar waveguide resonators. Our results demonstrate notable power and temperature dependence of dielectric loss, with the dielectric loss and quasiparticle loss exchanging dominance at around 4 K. These findings are consistent with TLS models and provide crucial insights for the design of superconducting devices operating at liquid helium temperatures.
\end{abstract}

\maketitle

The superconducting state significantly reduces conducting loss,  making other types of loss more prominent.  Among these, extrinsic dielectric loss due to two-level systems (TLSs) has been extensively studied  at millikelvin temperatures (1-400  mK), primarily driven by advancements in superconducting quantum devices and kinetic inductance detectors (KIDs) based on superconducting microwave resonators. This is because TLS loss plays a critical role in determining the coherent time of qubits and the sensitivity of the detectors \cite{mcrae2020materials}.In contrast, at liquid helium temperature (approximately 4 K), there has been surprisingly little experimental assessment of dielectric loss. This is despite the fact that various superconducting electronic devices operate at this temperature, including Josephson voltage standards \cite{hamilton1997josephson} , rapid single flux quantum logic \cite{chen1998superconductor,soloviev2017beyond}, hot electron bolometers\cite{maezawa2015application,shurakov2015superconducting}, and  superconductor-insulator-superconductor (SIS) heterodyne mixers covering wide frequency range from millimeter wave to terahertz frequencies \cite{mena2010design,mahieu2011alma,kerr2014development,belitsky2018alma,uzawa2021development}.Even at intermedium sub-kelvin temperatures (0.4-1  K), the TLS loss of amorphous dielectric material has not been fully assessed, although it may be relevant for certain applications, such as antenna-coupled transition edge sensors (TES)\cite{hubmayr2015feedhorn,hubmayr2016design}.

A conventional niobium SIS mixer circuit uses amorphous $\rm{SiO_2}$ as the dielectric layer for superconducting microstrip(MS) lines in tuning circuits and impedance transformers. $\rm{SiO_2}$ is chosen for these circuits due to its low permittivity ($\varepsilon_r\approx4$), in addition to superior chemical and mechanical stability. A low permittivity allows for a higher achievable characteristic impedance in microstrip lines, given the minimum strip width allowed in fabrication. This flexibility in selecting characteristic impedance is crucial for designing broadband tuning circuits and impedance matching networks between the low-impedance SIS junctions and the higher-impedance waveguides.

Conventionally designed SIS mixers usually contained electrically short transmission lines, typically less than one wavelength. As a result, dielectric loss was unlikely to cause noticeable degradation in the performance and thus received little attention. This is similar to the tolerance of quasiparticle loss at 4 K operation. However, advancements in SIS mixer technology have led to the use of larger-scale on-chip systems  for  compactness\cite{ar1998tunerless,koshelets2000integrated, vernik2005integrated}, multibeam\cite{shan2020compact,wenninger2023design}, and multi-band\cite{groppi2019first}, which involve significantly longer transmission lines. In these applications the accumulation of the losses from the dielectric layer may become pertinent.

\begin{figure}[tb]
\centering
\includegraphics[width=70mm,clip]{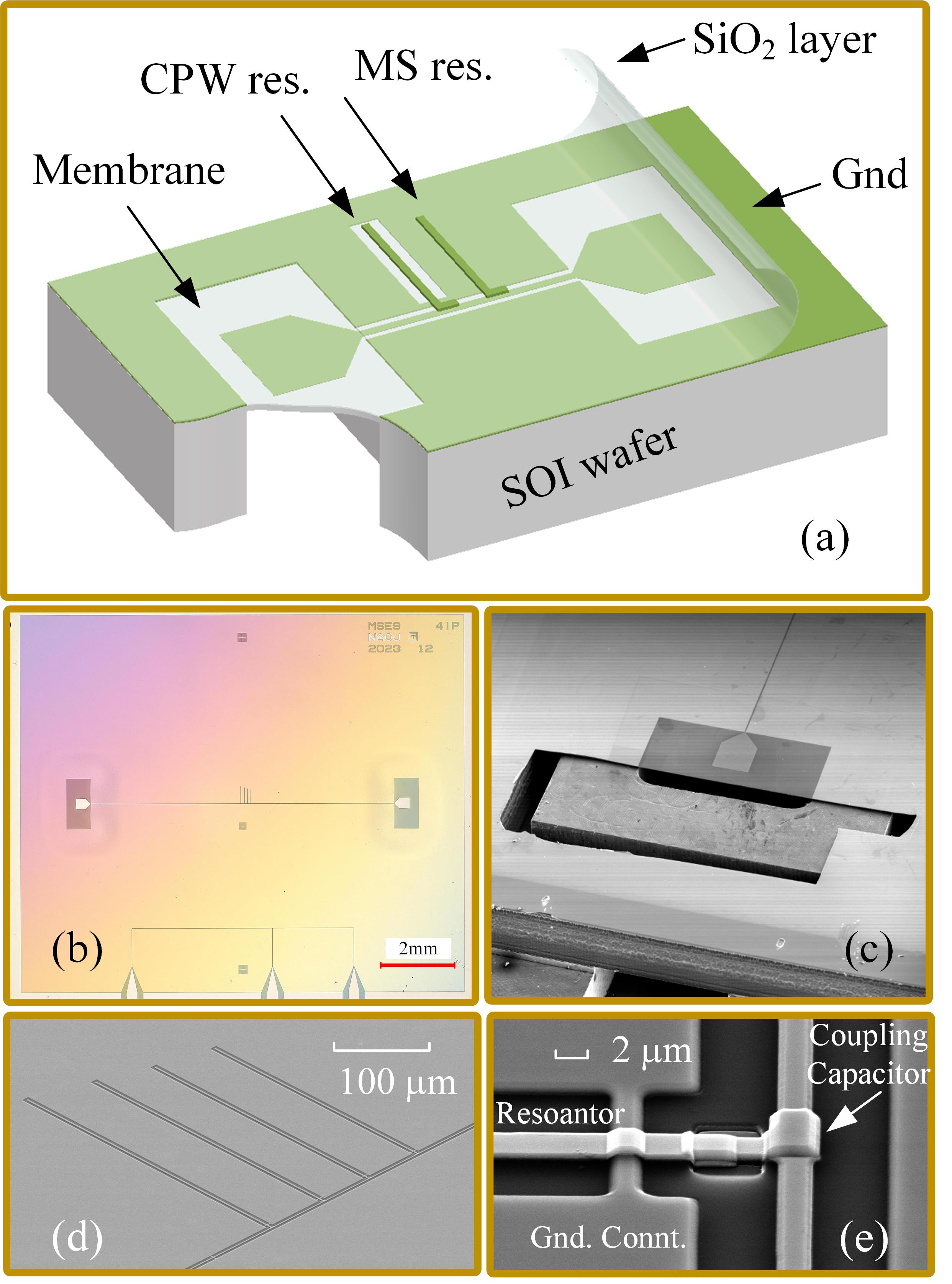}
\caption{(a) Schematic illustration of the on-chip two-port measurement network interfaced by a pair of silicon membrane-based waveguide-to-CPW transitions. A CPW resonator and an MS resonator capacitively couple to the readout CPW, which connects the two probes. An amorphous $\rm{SiO_2}$ dielectric layer is sandwiched between the conducting strips of the resonators and the ground plane. The mounting structure for the chip is detailed in the appendix. (b) photo of a chip. (c) SEM image of the silicon membrane which was broken for demonstration purpose. (d) SEM image of four CPW resonators. (e) SEM image of a coupling T-junction that couples a CPW resonator to the readout line. The ground planes on both sides of the CPW resonator are connected near the T-junction to avoid generating differential mode at the discontinuity.}
\label{fig1}
\end{figure}

There might be a mistaken belief that the TLS loss is negligible at 4 K due to the convergence of the occupations of the two states. However, thermalization depends on energy splitting, which has a broad distribution\cite{anderson1972anomalous,phillips1987two}. While TLS loss is not expected to be significant at microwave around 4 K, it can become important at millimeter and sub-millimeter wave lengths.  Indeed, if frequency and temperature increase by the same factor, the TLS loss does not change because it depends on the normalized temperature with respect to $\rm{\hbar\omega/k_B}$, appearing as a factor in the standard TLS loss model:
\begin{equation}\label{Eq1}
  \delta\left(T,\omega,P\right)=F\delta_{TLS}^0\left(\omega\right)\frac{tanh\frac{\hbar\omega}{2k_BT}}{\sqrt{1+\frac{P}{P_c}}}\,,
\end{equation}
where $\delta_{TLS}^0$ is TLS loss at 0 K and low power ($P\ll P_c$), $P_c$ is saturation power, and $\omega$, T, $k_B$ are angular frequency, temperature, and Boltzmann constant respectively. $F$ is the filling factor of the lossy dielectric material, which is close to 1 for an MS. $\delta_{TLS}^0$ , reflecting the density of state of TLSs, is smoothly dependent on frequency on the scale of $k_BT$\cite{anderson1972anomalous}.  This equation suggests that TLS loss at 160 GHz  and a temperature of  4 K  will be comparable to that at  4 GHz  and  100 mK.

Loss measurements of superconducting thin-film transmission lines at millimeter wave frequencies and 4 K have primarily relied on on-chip SIS detectors \cite{vayonakis2002millimeter,shan2024investigating}. This approach is favored because constructing a planar circuit-to-metal waveguide transition, which is necessary for network analyzer measurements, is challenging. However, the low saturation power (microwatt level) of SIS detectors limits the ability to identify the TLS nature from the power-dependence property, as these detectors saturate well below $P_c$. Although high-dynamic range measurements using network analyzers become less challenging at microwave frequencies, due to the use of coaxial cables to access planar circuits, TLS loss is heavily suppressed due to thermalization at 4 K. This suppression makes it difficult to distinguish TLS loss from the more significant quasiparticle loss. Notably, a network analyzer-based measurement of $\rm{SiO_2}$ loss was conducted at 70-75 GHz and 4.2 K using a fin-line microstrip-to-waveguide transition \cite{schubert2011microwave}. However, this study did not address the nature of the dielectric loss.

We measured $\rm{SiO_2}$ loss from the Q factor of half-wavelength niobium MS and coplanar waveguide (CPW) resonators, resonating in the frequency range between 130 - 170 GHz, and being fabricated on silicon chips. Each chip contains four resonators, spaced approximately 10 GHz apart, and coupled in parallel to a readout CPW, as shown in Fig. \ref{fig1}. The niobium films, deposited using DC magnetron sputtering, have a thickness of about 200 nm, sufficiently thicker than the penetration depth at 4 K. These films exhibit a $T_c$ of about 9 K and a residual resistivity of about $\rm{4.2\,\mu\Omega cm}$.  The amorphous $\rm{SiO_2}$ dielectric layer for the MS was deposited using plasma enhanced chemical vapor deposition (PECVD) or magnetron RF sputtering. Since the $\rm{SiO_2}$ layer covers the entire wafer above the niobium ground plane, it also lies between the strip and ground plane of the CPW resonators, resulting a filling factor $F\approx0.17$. The details of resonator designs can be found in the appendix. The chips containing resonators were cooled down in a cryostat by using a close-cycled GM cooler to a minimum stage temperature approximately  3.3 K, monitored with a calibrated temperature sensor. The stage can be heated up to the transition temperature of the niobium using a heater.

The measurement was carried out with a millimeter wave network analyzer. The silicon membrane-based waveguide-to-CPW transition embedded in the silicon wafer enable the transition from the CPW to the external WR-6 waveguide for connecting the network analyzer. Despite of a high permittivity of silicon, due to the very short electric length inside the 6-$\rm{\mu m}$ thick membrane, the frequency-dependence in the transmission characteristics can be well suppressed. In consequence, this waveguide-to-CPW transition has a high coupling efficiency with a return loss less than -15 dB over a broad bandwidth covering  125 - 211 GHz  \cite{masukura2023silicon}. A pair of such transitions are deployed back-to-back on the silicon chip, constructing a versatile and general-purpose test platform for the assessment of two-port planar circuit components.  Details of the measurement setup can be found in the appendix.

\begin{figure}[tb]
\centering
\includegraphics[width=70mm,clip]{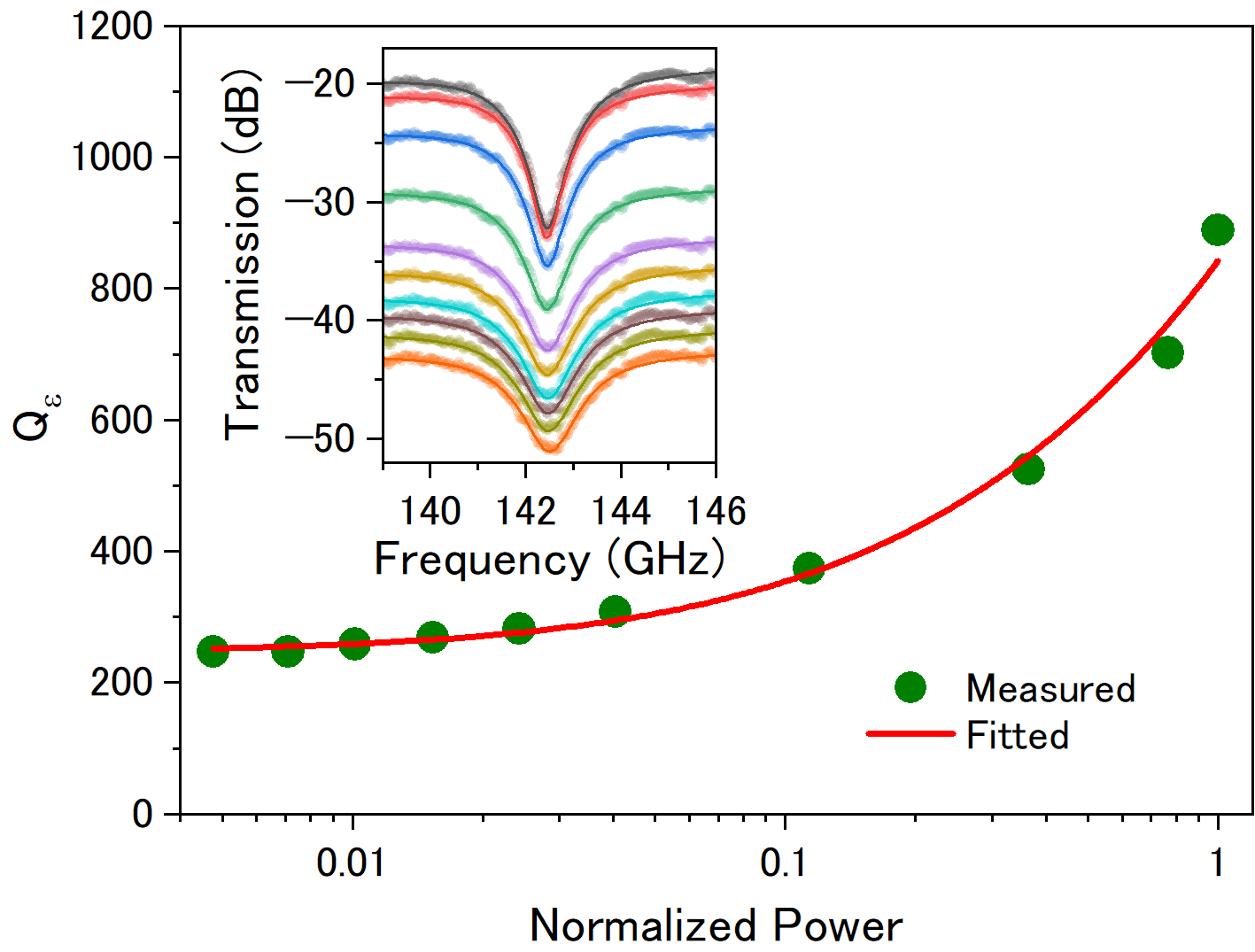}
\caption{Dielectric Q factor ($Q_\varepsilon$) of an MS resonator measured at  3.3  K  against signal power(dots), which is normalized to approximately 3 $\rm{\mu W}$ at zero attenuation. The $\rm{SiO_2}$ layer of this sample was deposited using PECVD.  The inset shows the measured and fitted resonance curves of the resonator at various power levels. }
\label{fig2}
\end{figure}

Calibrating the network analyzer at cryogenic temperatures is usually more demanding than at room temperature. However, cryogenic calibration can be avoided here for two reasons.  First, only the amplitude of transmission response is needed. Second, the amplitude of standing waves caused by discontinuities in the measurement path external to the network analyzer was carefully minimized, achieving a low ripple less than $\rm{0.25\,dB}$ in RMS within the 130-170 GHz  band. Since standing waves of this level did not significantly affect the resonance curve fitting in the data analysis, it is sufficient to calibrate the network analyzer only at room temperature external to the cryostat.

\begin{figure*}[tb]
\centering
\includegraphics[width=160mm,clip]{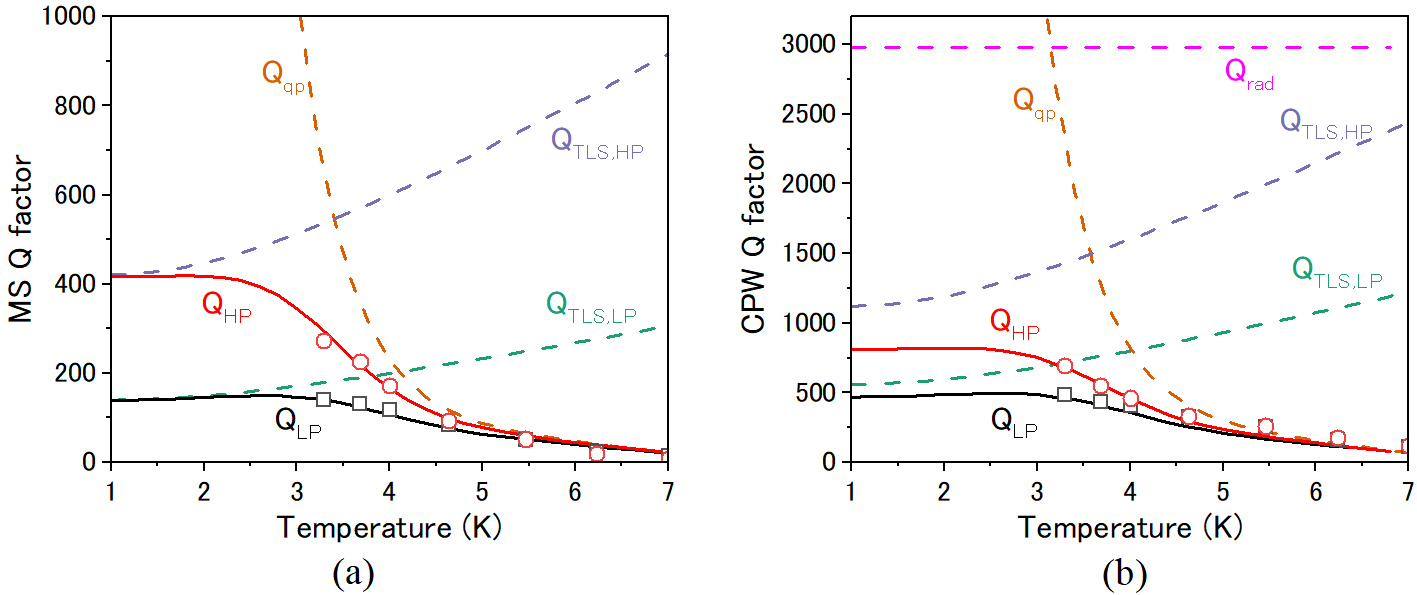}
\caption{(a) Measured (dots) and modeled (solid lines) intrinsic Q factor of an MS resonator and (b) a CPW resonator with respect to ambient temperature under high power ($Q_{HP}$ at 3 $\rm{\mu W}$) and low power ($Q_{LP}$). The overall Q is decomposed into compositions of quasiparticle ($Q_{qp}$), TLS ($Q_{TLS}$) and radiation ($Q_{rad}$) losses, depicted in dashed lines.  }
\label{fig3}
\end{figure*}

The intrinsic Q factor of a resonator is obtained by fitting the measured resonance curve with $\phi$-rotation model\cite{gao2008physics}. To extract the dielectric loss from the overall loss, the quasiparticle loss, which is significant at liquid helium temperature, should be determined first. The Q factor of the resonators measured at relatively high temperature (>5 K, or 0.6 $T_c$) was used in quantifying the quasiparticle loss because model predictions safely indicate that quasiparticle loss is dominant at these temperatures.  that quasiparticle loss is dominant at these temperatures. We applied a transmission line model to align the calculated loss with the measured one at this temperature range by tuning material parameters of niobium, as detailed in the appendix. Then, the quasi-particle loss at lower temperature was calculated by using this refined model and subtracted from the overall loss to obtain the dielectric loss. Additionally, because TLS loss is strongly dependent on frequency according to the factor $tanh\left(\hbar\omega/k_BT\right)$, we also fabricated MS resonators resonating at approximately  5 GHz  using the same fabrication process as for millimeter resonators. This approach provides an extended scope to identify the nature of the dielectric loss.

TLS losses are distinctive due to their unique power-dependence feature. At temperatures lower than 5 K, the power-dependences of the intrinsic Q factor of both MS and CPW resonators were clearly observed.  An example of an MS resonator measured at 3.3 K is shown in the inset of Fig. \ref{fig2}.  After removing the quasiparticle loss, the dielectric Q factors are plotted against signal power in Fig.\ref{fig2}. The data align well with the standard TLS model shown in Eq. \ref{Eq1}, with an intrinsic TLS loss $\delta_{TLS}^0\approx6.7\times{10}^{-3}$ for $\rm{SiO_2}$ deposited by using PECVD and a critical power $P_c\approx\rm{0.3\mu\,W}$, corresponding to a field of $E_c\approx\rm{0.5\,MV/m}$.  In the fitting of TLS loss, we have assumed that the losses other than the quasiparticle loss and the TLS loss can be neglected. This assumption is justified for two reasons: First, any frequency-independent loss comparable to the two major losses can be ruled out, as the microwave version of the resonators at about 5 GHz show significantly lower loss by a factor of 20 at 4 K; and second, the only known frequency-dependent radiation loss is not relevant in MS.

The temperature dependence of the intrinsic Q factor of MS and CPW resonators was measured under both weak ($P\ll P_c$) and strong power (about $\rm{3\,\mu W}$). The measured temperature dependence closely matches the model, with $\delta_{TLS}^0$ and $P_c$ for TLS loss and $\sigma_n$ for quasiparticle loss determined experimentally. Examples for an MS resonator and a CPW resonator with PECVD-deposited $\rm{SiO_2}$ are shown in Fig. \ref{fig3}.  The breakdown of the intrinsic Q factor into the major compositions was performed using the previously discussed quasiparticle and TLS models, as well as a radiation loss model\cite{kasilingam1983surface,rutledge1983integrated} for the CPW resonator. The radiation loss of the CPW, with a configuration of slot/strip/slot=$\rm{4\,\mu m/2\,\mu m/4\,\mu m}$ on silicon substrate, is predicted to be insignificant at this frequency band compared to other losses.  One important observation is that the quasiparticle loss of niobium and the dielectric loss will become comparable and exchange their leading roles in the temperature range between 3 to 4 K. Even if the TLS noise can be reduced to some extent by improving fabrication techniques, the crossover still occurs in this temperature range because, there,  the quasiparticle loss rapidly reduces with temperature.

The power dependence of the intrinsic Q factor of the CPW resonator is weaker than that of MS resonators, as shown by comparing Fig. \ref{fig3}(a) with (b). It is because the TLS loss is reduced by the filling factor. We applied buffered HF etch to the chips containing CPW resonators to remove the $\rm{SiO_2}$ in the slots and partly under the strips. Although the remove of $\rm{SiO_2}$ caused an upward frequency shift of the resonator by about 20\%, some resonators remained still within the measurement frequency range, allowing us to confirm that  the power dependence of the resonance curves became unobservable. For MS resonators, however, the resonance frequencies became too high and fell outside the measurable frequency range.

Resonators with sputtered $\rm{SiO_2}$ layer fabricated using a magnetron RF plasma system were also measured. Measured results show $\delta_{TLS}^0\approx3\times{10}^{-3}$, which is factor of 2 smaller than those deposited with PECVD. These results align with those measured by others at similar frequency range: a tangential loss of $5.3 \times 10^{-3}$ for sputtered $\rm{SiO_2}$ at frequencies between  75-100 GHz and at  4.2  K  were reported by Vayonakis et al.\cite{vayonakis2002millimeter}, and losses ranging from $0.5-2 \times 10^{-3}$ at approximately  100 GHz  and  30 mK  were observed by Gao et al.\cite{gao2009measurement}, with both measurements conducted using superconducting MS resonators. At microwave frequencies, more measurement data\cite{mcrae2020materials} show a wider range of $\delta_{TLS}^0$ from $3\times 10^{-4}$ to $7 \times 10^{-3}$.

\begin{figure}[tb]
\centering
\includegraphics[width=70mm,clip]{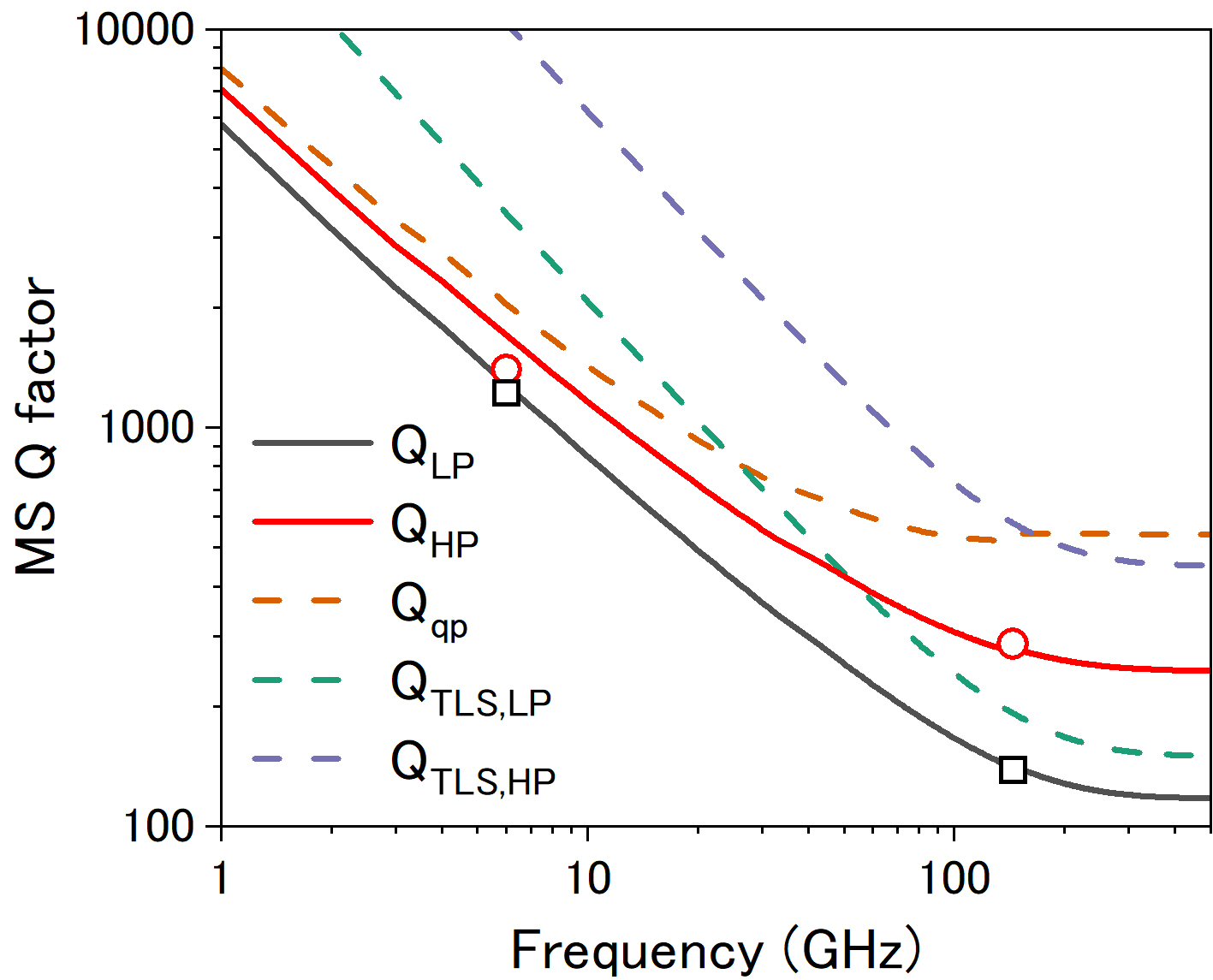}
\caption{Measured intrinsic Q factors (dots) of microwave and millimeter-wave MS resonators at 3.3 K under strong power (about $\rm{3\,\mu W}$) and low power ($P\ll P_c$). Solid lines represent the model-predicted Q factors as a function of frequency up to sub-millimeter range. Dashed lines depict the decomposed TLS noises (under high power and low power) and quasiparticle loss. }
\label{fig4}
\end{figure}

At  5 GHz, the TLS loss is predicted to be significantly smaller than that at  140 GHz, becoming less significant than the quasiparticle loss. This makes identification of TLS loss at  4 K  difficult at microwave. In our measurement, the intrinsic Q factor of MS resonators with $\rm{SiO_2}$ deposited with PECVD  were found to be about $1.4\times10^3$. After correcting for the quasiparticle loss, the weak-field dielectric loss is about $2.5\times{10}^{-4}$, corresponding to $\delta_{TLS}^0\approx7\times{10}^{-3}$, which is very close to the intrinsic TLS loss obtained at 2 mm wavelength. We observed a weak power dependence of the intrinsic Q factor. When the input power, corrected for the transmission loss, was increased to $3\,\mu W$, which is the "high power" applied in mm measurement, the intrinsic Q factor increased by about 10\% with respect to the weak-field value, as shown in Fig. \ref{fig4}. The increase is about 40\% of  that predicted by the model, which assumes that $P_c$ is independent of frequency. This suggests that $P_c$ at about 5 GHz is as large approximately twice the value at about  140 GHz.

Assuming that the density of states and saturation power of TLSs, which are expected to be smoothly dependent frequency, remain constant in a frequency range from about 140 GHz to the gap frequency of niobium, we can make meaningful predictions for the niobium MS loss at sub-millimeter range, as shown in Fig. \ref{fig4}. It is found that TLS loss and quasiparticle loss tend to level off as frequency increases toward the sub-millimeter range. Therefore, their relative values will not change significantly compared to those at 140 GHz. This is a favorable signal for the sub-millimeter applications at 4 K using MS lines. For superconducting CPWs, the radiation loss may become dominate at submillimeter waves\cite{grischkowsky1987electromagnetic,Hahnle2020}.

In conclusion, we have measured the  power-dependent intrinsic Q factor of both $\rm{SiO_2}$-loaded MS and CPW resonators at  2  mm wavelength and found that they can be accurately modeled using the TLS model. At liquid helium temperatures, the TLS loss of amorphous $\rm{SiO_2}$ may becomes a dominant factor in planar superconducting transmission lines, in particular MSs,  operating at millimeter and sub-millimeter wavelengths. This study highlights the significance of TLS loss at 4 K,  where a crossover  between quasiparticle loss and TLS loss occurs, providing crucial insights for the design of superconducting devices operating in this temperature range.

\begin{acknowledgments}
The work is partly supported by the Japan Society for the Promotion of Science(JSPS) KAKENHI under Grant Number 23K20871.
\end{acknowledgments}

\section*{Supplementary Material}

See the supplementary material for detailed information, including measurement setup, resonator design and fabrication, and correction for quasiparticle loss.

\subsection{Measurement setup}

\begin{figure}[h]
\centering
\includegraphics[width=80mm,clip]{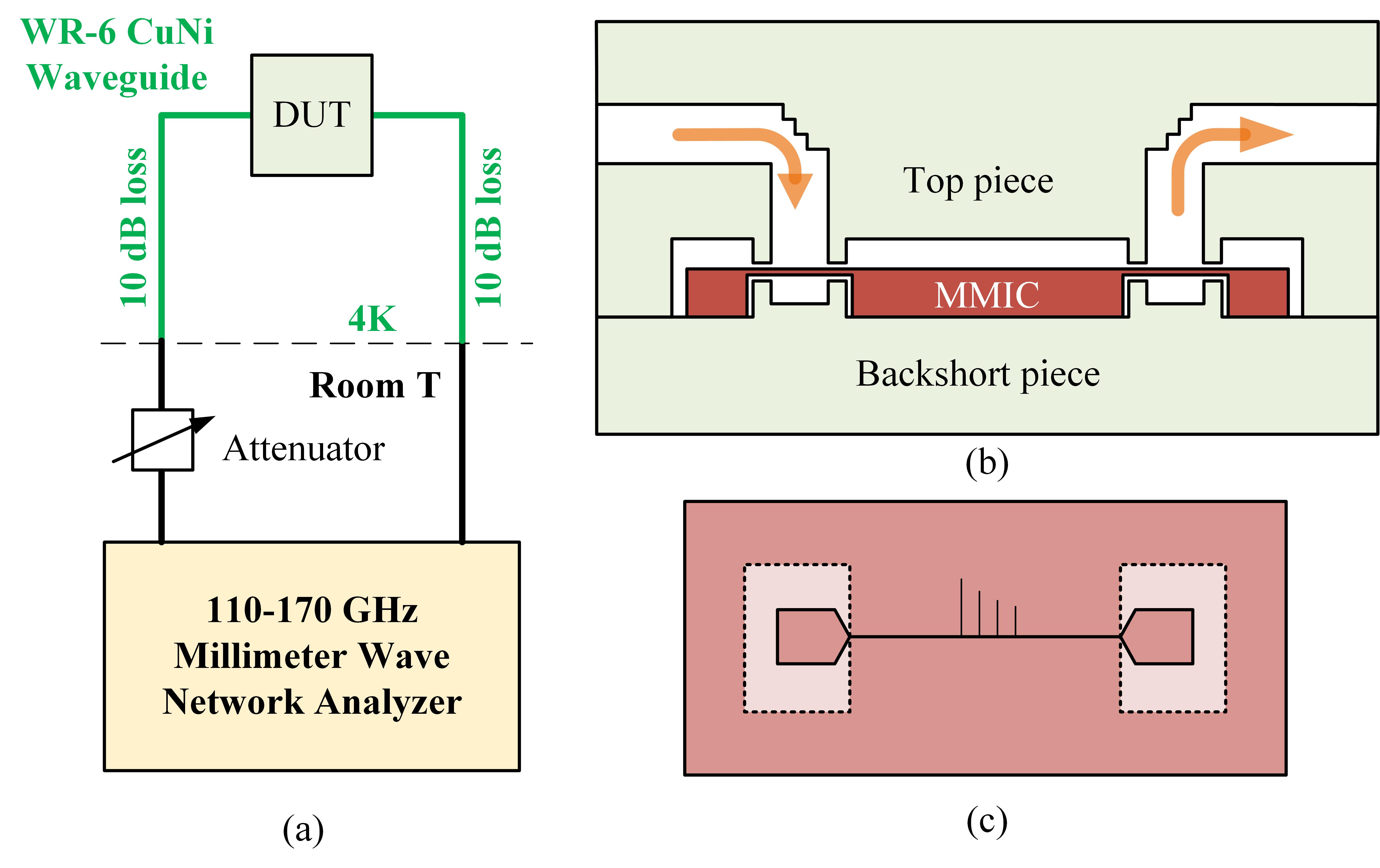}
\caption{Setup for millimeter wave resonator measurement. (a) Diagram of the measurement system. (b) Schematic structure of the enclosure of the MMIC containing resonators. (c) Schematic top-view of the MMIC. }
\label{fig5}
\end{figure}

The measurement setup, as shown in Fig. \ref{fig5}(a) employs an Agilent E8362C vector network analyzer (NA) and an N5260A millimeter head controller together with OML V06VNA2 millimeter wave VNA extenders. A D-band variable attenuator is used to vary the signal power. The millimeter wave monolithic integrated circuit (MMIC) containing resonators is mounted in an aluminum enclosure with build-in waveguides and standard waveguide interfaces. This unit, referred to as device-under-test (DUT), is thermally anchored to the 4-K cold stage and connected to the NA with thermal-isolating CuNi WR-6 waveguides on each side, with each having an attenuation of about 10 dB. The power measured at the output of the  millimeter wave transducer is approximately 30 $\rm{\mu W}$ using an mm-wave powermeter Erikson PM4. The structure of the DUT is illustrated in Fig. \ref{fig5}(b) and \ref{fig5}(c). Two embedded waveguide H-bends are designed to have a passband wider than the silicon membrane-based waveguide-to-CPW transitions. The measured transmission performance of this module is detailed in a previous study\cite{masukura2023silicon}.

\subsection{Resonator design and fabrication}
\begin{figure}[h]
\centering
\includegraphics[width=60mm,clip]{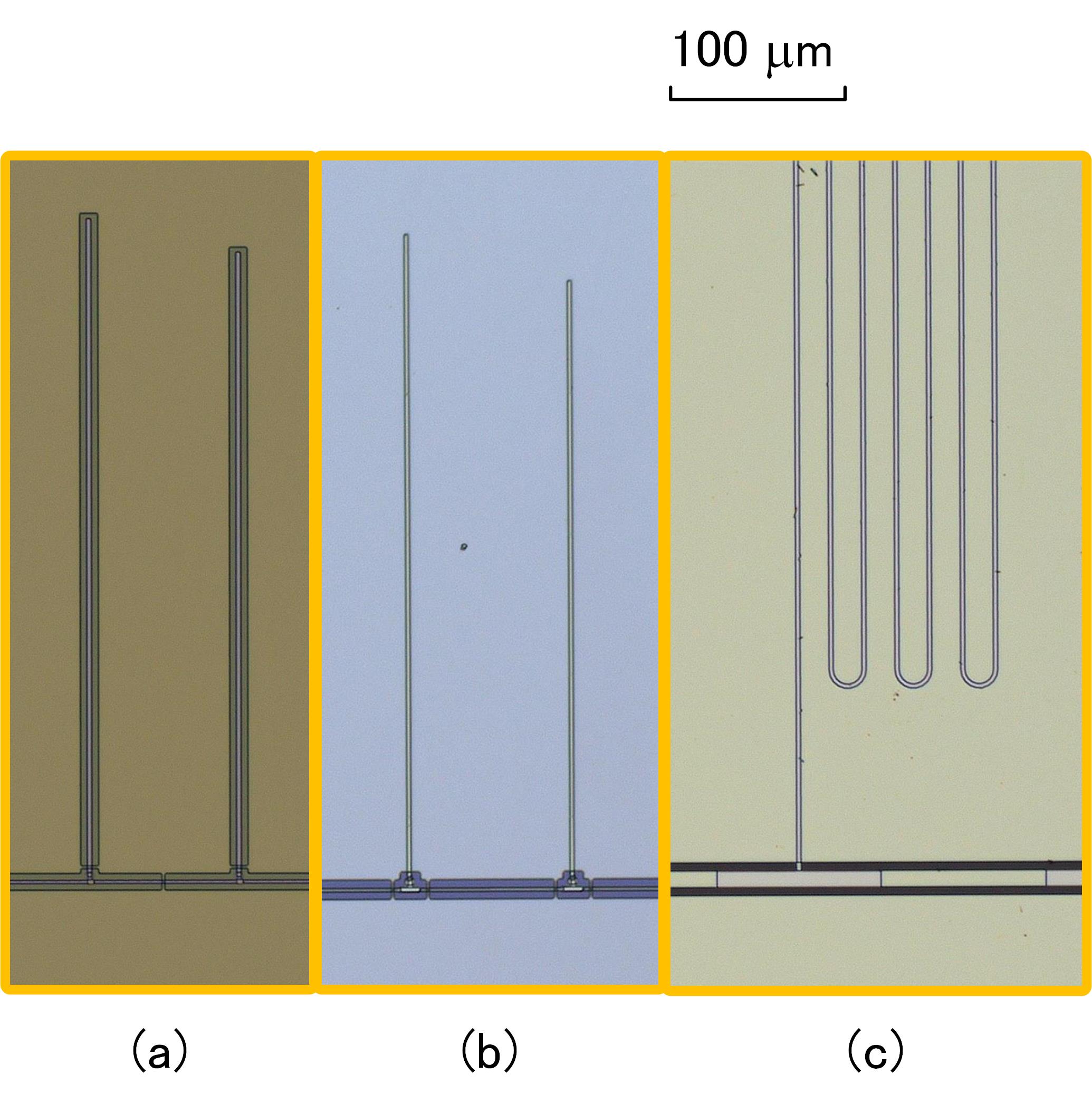}
\caption{Microscopic images of (a) millimeter wave CPW resonators, (b) millimeter wave MS resonators, and (c) a microwave MS resonators. }
\label{fig6}
\end{figure}

The ground plane and the conducting strips for MSs and CPWs were made from niobium thin films with thickness of about  200 nm. The conducting strip width of MSs and CPWs are 3 $\rm{\mu m}$ and 2 $\rm{\mu m}$ respectively. The configurations of CPWs for readout and resonators are $\rm{4\,\mu m/2\,\mu m/4\,\mu m}$ (slot/strip/slot) for mm-wave resonators and $\rm{5\,\mu m/10\,\mu m/5\,\mu m}$ for the readout CPW at microwave frequencies. Microscopic images of the resonators are shown in Fig.\ref{fig6}. The $\rm{SiO_2}$ is 300 nm thick for mm-wave resonators and 200 nm thick for microwave resonators; this layer completely covers all the wafer above the ground patterns. The resonators couple signals from the readout line using parallel plate capacitors with the same $\rm{SiO_2}$ layer as dielectric material. The bottom plate of each capacitor, a patch isolated from the ground plane, is connected to the center strip of the resonator through a conducting via hole cross the $\rm{SiO_2}$ layer. The symmetry of the CPW is disrupt at the coupling T-junctions because resonators are asymmetrically coupled on one side. The broken symmetry may stimulate the CPW odd mode, leading to a reduction in intrinsic Q.  To avoid the odd mode generation, bridges were used to connect the ground planes of the readout CPW around the T-junctions. The bridges were made on the ground plane to ease the fabrication, rather than using conventional air-bridges. They cause negligible reflection because the bridges are electrically very short.

Niobium was deposited using DC magnetron sputtering at a substrate temperature $\rm{20\, ^oC}$. $T_c$ was measured about 8.9 K. The film exhibited compressive stress of about $\rm{-0.65\,GPa}$, and the resistivity at room temperature is about $\rm{18.6 \,\mu\Omega cm}$. The residual resistance ratio was measured to be $4.5$, resulting in a residual resistivity approximately $\rm{4.2 \,\mu\Omega cm}$ at the temperature immediately above the superconducting transition. $\rm{SiO_2}$ was deposited in a PECVD system, Samco PD-200STL, using tetraethyl orthosilicate (TEOS) as the precursor. The deposition was carried out in an oxygen-rich atmosphere with flow rate of TEOS and $\rm{O_2}$ at 15 and 680 sccm respectively, and a pressure of 40 Pa. The reactor temperature was set to be $\rm{100\,^oC}$, and the plasma power was 350 W. The resulting film exhibited a compressive stress of about $-0.32$ GPa. For some devices, RF magnetron sputter (ULVAC MB951011) was used. The film was sputtered in an argon plus 10\% oxygen atmosphere at $3.6$ Pa and 400 W at room temperature. The film was nearly stress-free.

\subsection{Correction for quasiparticle loss}

To extract the dielectric loss from the overall loss of a resonator implied by intrinsic Q factor, it is necessary to correctly determine the quasiparticle loss. The quasiparticle loss of an MS was calculated by combining the Mattias-Bardeen theory \cite{mattis1958theory} and a conformal mapping technique \cite{chang1976analytical,gevorgian1995cad}, which takes into account the geometric configuration of the MS. Our first attempt to calculate the quasiparticle loss with using the measured niobium residue resistivity $\rm{4.2\,\mu\Omega cm}$ was inconsistent with the measurement results. The calculated quasiparticle loss at $\rm{T>5\,K}$, where quasiparticle loss is dominant, was about half as large as measured values. The analytical transmission line model is unlikely to have caused the discrepancy because numerical simulations using EM field simulator (HFSS\cite{hfss}) and complex surface impedances \cite{Kerr1999} resulted in very similar results to the analytical ones. We tried tuning the niobium parameters, including gap voltage (nominally  2.8 mV), $T_c$ (8.9 K measured), conductivity, and electron density ($\rm{5.6\times10^{28}\,m^{-3}}$) to make the calculated results approach the measured ones and found that conductivity is the most sensitive and likely the responsible parameter. The inconsistency can be alleviated if the niobium residue resistivity is set to $\rm{12.5 \,\mu\Omega cm}$, about three times the value  measured at DC. This implies that the niobium film has inhomogeneous quality with a degraded epilayer contributing most to RF loss, though this degradation cannot be reflected in the DC measurement. This phenomenon is also observed by Lodewijk etc.\cite{lodewijk2007optimizing}, and is supported by the fact that niobium film quality is strongly dependent on thickness, showing a gradually development of texture in the initial layer\cite{minhaj1994thickness,Lacquaniti1995,chandrasekaran2021nb}. The effective residual resistivity was applied in the theoretical model to separate the quasiparticle loss and the dielectric loss in the data analysis.

It is worth noting that even though there is some uncertainty in the determination of quasiparticle loss, it does not significantly affect the extraction of dielectric loss at 4 K and lower temperatures. This is because the quasiparticle loss, calculated with either nominal or effective residual conductivity, becomes relatively unimportant compared to the dielectric loss, and does not qualitatively affect the conclusion. By using the effective residual resistivity, the consistency between the model and the measurement is achieved over a broad temperature range. The good match between the simulated and measured temperature dependance further justifies the use of the effective conductivity.

%\nocite{*}
\bibliography{TLSLetter}% Produces the bibliography via BibTeX.

\end{document}